\documentclass[showpacs, preprintnumbers, pra, superscriptaddress, floatfix, twocolumn, longbibliography]{revtex4-1}

\usepackage{latexsym}
\usepackage{natbib}
\usepackage[utf8]{inputenc}
\usepackage{graphicx}% Include figure files
\usepackage{dcolumn}% Align table columns on decimal point
\usepackage[usenames, dvipsnames]{color}
\usepackage{bm}% bold math
\usepackage{color}
\usepackage{hyperref}% add hypertext capabilities
\newcommand{\mnbite}{\ensuremath{\text{MnBi}_4\text{Te}_7}}
\newcommand{\mnbi}{Mn$_{0.75}$Bi$_{4.25}$Te$_7$}
\newcommand{\mnbiv}{Mn$_{0.50}\Box_{0.25}$Bi$_{4.25}$Te$_7$}

\begin{document}

\title{Topological electronic structure and intrinsic magnetization in MnBi$_4$Te$_7$: \\ a Bi$_2$Te$_3$-derivative with a periodic Mn sublattice}

\author{Raphael C. Vidal}%
\affiliation{Experimental Physics VII, Universität Würzburg, D-97074 Würzburg, Germany}%
\affiliation{Würzburg-Dresden Cluster of Excellence ct.qmat}%
\affiliation{These authors contributed equally to this work}%
\author{Alexander Zeugner}%
\affiliation{Faculty of Chemistry and Food Chemistry, Technische Universität Dresden, D-01062 Dresden, Germany}%
\affiliation{These authors contributed equally to this work}%
\author{Jorge I. Facio}%
\affiliation{Leibniz IFW Dresden, Helmholtzstr. 20, D-01069 Dresden, Germany}%
\affiliation{These authors contributed equally to this work}
\author{Rajyavardhan Ray}%
\affiliation{Leibniz IFW Dresden, Helmholtzstr. 20, D-01069 Dresden, Germany}%
\author{M. Hossein Haghighi}%
\affiliation{Leibniz IFW Dresden, Helmholtzstr. 20, D-01069 Dresden, Germany}%
\author{Anja U. B. Wolter}%
\affiliation{Leibniz IFW Dresden, Helmholtzstr. 20, D-01069 Dresden, Germany}%
\author{Laura T. Corredor Bohorquez}%
\affiliation{Leibniz IFW Dresden, Helmholtzstr. 20, D-01069 Dresden, Germany}%
\author{Federico Caglieris}%
\affiliation{Leibniz IFW Dresden, Helmholtzstr. 20, D-01069 Dresden, Germany}%
\author{Simon Moser}%
\affiliation{Advanced Light Source, Lawrence Berkeley National Laboratory, Berkeley, CA 94720, USA}%
\author{Tim Figgemeier}%
\affiliation{Experimental Physics VII, Universität Würzburg, D-97074 Würzburg, Germany}%
\affiliation{Würzburg-Dresden Cluster of Excellence ct.qmat}%
\author{Thiago R. F. Peixoto}%
\affiliation{Experimental Physics VII, Universität Würzburg, D-97074 Würzburg, Germany}%
\affiliation{Würzburg-Dresden Cluster of Excellence ct.qmat}%
\author{Hari Babu Vasili}%
\affiliation{ALBA Synchrotron Light Source, E-08290 Cerdanyola del Valles, Spain}%
\author{Manuel Valvidares}%
\affiliation{ALBA Synchrotron Light Source, E-08290 Cerdanyola del Valles, Spain}%
\author{Sungwon Jung}%
\affiliation{Diamond Light Source, Harwell Campus, Didcot OX11 0DE, United Kingdom}%
\author{Cephise Cacho}%
\affiliation{Diamond Light Source, Harwell Campus, Didcot OX11 0DE, United Kingdom}%
\author{Alexey Alfonsov}%
\affiliation{Leibniz IFW Dresden, Helmholtzstr. 20, D-01069 Dresden, Germany}%
\author{Kavita Mehlawat}%
\affiliation{Leibniz IFW Dresden, Helmholtzstr. 20, D-01069 Dresden, Germany}%
\affiliation{Würzburg-Dresden Cluster of Excellence ct.qmat}%
\author{Vladislav Kataev}%
\affiliation{Leibniz IFW Dresden, Helmholtzstr. 20, D-01069 Dresden, Germany}%
\affiliation{Würzburg-Dresden Cluster of Excellence ct.qmat}%
\author{Christian Hess}%
\affiliation{Leibniz IFW Dresden, Helmholtzstr. 20, D-01069 Dresden, Germany}%
\author{Manuel Richter}%
\affiliation{Leibniz IFW Dresden, Helmholtzstr. 20, D-01069 Dresden, Germany}%
\affiliation{Dresden Center for Computational Materials Science (DCMS), Technische Universität Dresden, D-01062 Dresden, Germany}%
\author{Bernd Büchner}%
\affiliation{Leibniz IFW Dresden, Helmholtzstr. 20, D-01069 Dresden, Germany}%
\affiliation{Würzburg-Dresden Cluster of Excellence ct.qmat}%
\affiliation{Faculty of Physics, Technische Universität Dresden, D-01062 Dresden, Germany}%
\author{Jeroen van den Brink}%
\affiliation{Leibniz IFW Dresden, Helmholtzstr. 20, D-01069 Dresden, Germany}%
\affiliation{Würzburg-Dresden Cluster of Excellence ct.qmat}%
\affiliation{Faculty of Physics, Technische Universität Dresden, D-01062 Dresden, Germany}%
\author{Michael Ruck}%
\affiliation{Faculty of Chemistry and Food Chemistry, Technische Universität Dresden, D-01062 Dresden, Germany}%
\affiliation{Würzburg-Dresden Cluster of Excellence ct.qmat}
\affiliation{Max Planck Institute for Chemical Physics of Solids, D-01187 Dresden, Germany}
\author{Friedrich Reinert}%
\affiliation{Experimental Physics VII, Universität Würzburg, D-97074 Würzburg, Germany}%
\affiliation{Würzburg-Dresden Cluster of Excellence ct.qmat}%
\author{Hendrik Bentmann}%
\email{hendrik.bentmann@physik.uni-wuerzburg.de}
\affiliation{Experimental Physics VII, Universität Würzburg, D-97074 Würzburg, Germany}%
\affiliation{Würzburg-Dresden Cluster of Excellence ct.qmat}%
\author{Anna Isaeva}%
\email{anna.isaeva@tu-dresden.de}
\affiliation{Leibniz IFW Dresden, Helmholtzstr. 20, D-01069 Dresden, Germany}%
\affiliation{Würzburg-Dresden Cluster of Excellence ct.qmat}%
\affiliation{Faculty of Physics, Technische Universität Dresden, D-01062 Dresden, Germany}%

\date{\today}

\begin{abstract}
Combinations of non-trivial band topology and long-range magnetic order hold promise for realizations of novel spintronic phenomena, such as the quantum anomalous Hall effect and the topological magnetoelectric effect. Following theoretical advances, material candidates are emerging. 
Yet, so far a compound that combines a band-inverted electronic structure with an intrinsic net magnetization remains unrealized. 
MnBi$_2$Te$_4$ has been established as the first antiferromagnetic topological insulator and constitutes the progenitor of a modular (Bi$_2$Te$_3$)$_n$(MnBi$_2$Te$_4$) series. Here, for $n = 1$, we confirm a non-stoichiometric composition proximate to 
\mnbite . We establish an antiferromagnetic state below $13$~K followed by a state with net magnetization and ferromagnetic-like hysteresis below $5$~K. Angle-resolved photoemission experiments and density-functional calculations reveal a topologically non-trivial surface state on the MnBi$_4$Te$_7$($0001$) surface, analogous to the non-magnetic parent compound Bi$_2$Te$_3$. 
Our results establish MnBi$_4$Te$_7$ as the first band-inverted compound with intrinsic net magnetization providing a versatile platform for the realization of magnetic topological states of matter.
\end{abstract}

\pacs{Valid PACS appear here}
\maketitle

Soon after the discovery of topological insulators (TIs) a decade ago \cite{hasan2010rmp}, the role of magnetism and its potential to modify the electronic topology emerged as a central issue in the field of topological materials. Magnetic degrees of freedom provide a powerful means of tuning the decisive characteristic of any topological system: its symmetry. By now it is recognized that the interplay between magnetic order and electronic topology offers a rich playground for the realization of exotic topological states of matter, such as the quantum anomalous Hall state \cite{2,3}, the axion insulator state \cite{4,5,6}, and magnetic Weyl and nodal line semimetals \cite{7,8,9,liu2019magnetic,ghimire2019creating}, enabling in turn different routes to spintronic applications \cite{10,11,12}.

The non-trivial topology in paradigmatic TIs like Bi$_2$Te$_3$ is a result of band inversion driven by strong spin-orbit interaction \cite{13,14}. Until recently, the interplay with magnetism in topological insulators has been mostly explored by extrinsic methods, such as doping a known TI with magnetic impurities~\cite{13} or interfacing it with ferromagnets~\cite{15}. Numerous derivatives of Bi$_2$Te$_3$ doped by transition metals have been explored over the last years \cite{2,3,16}, but the magnetically-active atoms did not form a periodic crystal sublattice. 
Initiated by works on epitaxial MnBi$_2$Se$_4$ layers \cite{hirahara2017large,hagmann2017molecular}, the compound MnBi$_2$Te$_4$ has recently arisen as the first derivative of Bi$_2$Te$_3$ that hosts structurally and magnetically ordered Mn atoms on well-defined crystallographic sites \cite{17,18,19,20}. The emergence of an antiferromagnetic TI state in MnBi$_2$Te$_4$ is now being broadly scrutinized by theoretical and experimental methods \cite{6,19,21,22,23,24,25,26}. 
At the same time, MnBi$_2$Te$_4$ provides the first example of an intrinsic magnetic TI \cite{19,24,26,27}.

In this joint experimental and theoretical work we establish another ternary manganese-bismuth telluride, MnBi$_4$Te$_7$, i.e. the ($n = 1$)-member of a modular (Bi$_2$Te$_3$)$_n$(MnBi$_2$Te$_4$) series, as the first instance of a compound that features, both, an inverted electronic band structure and an intrinsic net magnetization. Moreover, we observe several competing magnetic states in MnBi$_4$Te$_7$ which, in combination with the presence of topological surface states, could provide a versatile platform for tunability between different topological regimes. 

\section{Crystal growth and structure}

\begin{figure}
  \center
  \includegraphics[width = 8.5cm]{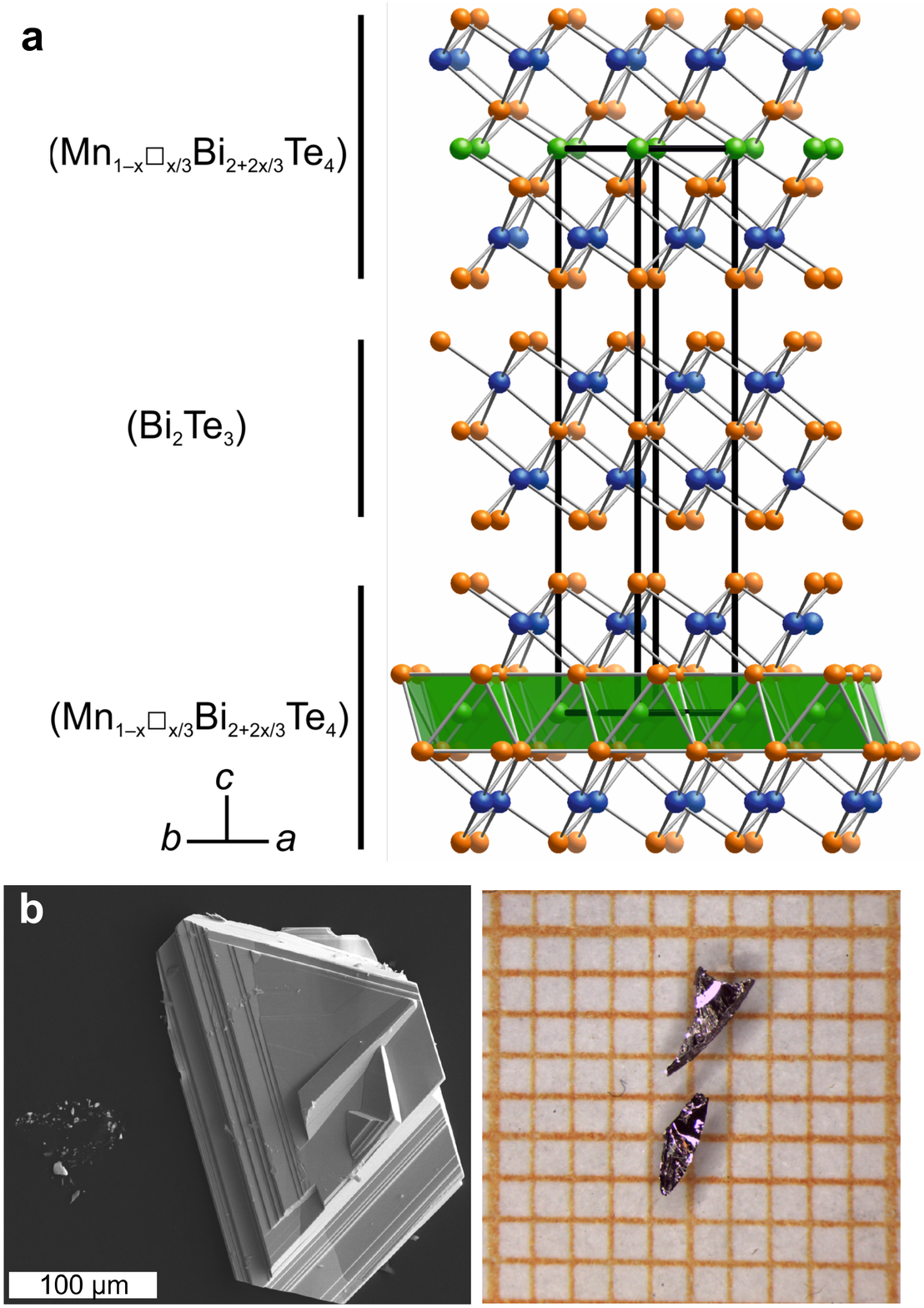}
  \caption{\label{fig:Mn147struct}  \textbf{a}, Crystal structure of Mn$_{1-x}$Bi$_{4+2x/3}$Te$_7$ (GeBi$_4$Te$_7$ structure type) with alternating (Bi$_2$Te$_3$) and (Mn$_{1-x}\Box_{x/3}$Bi$_{2+2x/3}$Te$_4$) blocks. Mn atoms are shown in green, Bi -- in blue, Te -- in orange. \textbf{b}, As-grown crystals with the experimental composition (EDX) Mn$_{0.8(1)}$Bi$_{4.3(1)}$Te$_7$.}
\end{figure}

 Recent systematic, synthetic explorations of the Mn$-$Bi$-$Te system have revealed new compounds that are ordered (Bi$_2$Te$_3$)$_n$(MnBi$_2$Te$_4$) ($n = 1, 2$) modular stackings of quintuple (Bi$_2$Te$_3$) and septuple (MnBi$_2$Te$_4$) blocks (Fig.~\ref{fig:Mn147struct}a) \cite{26,27}. In this section, we report a new crystal-growth protocol for $n = 1$, assisted by our preceding thermochemical studies~\cite{27}, and the structure elucidation by single-crystal X-ray diffraction (SCXRD). In Ref.~\cite{27} we developed robust synthetic protocols for phase-pure powders of $n = 1$ and $n = 2$ members based on differential scanning calorimetry. For all members of the (Bi$_2$Te$_3$)$_n$(MnBi$_2$Te$_4$) ($n = 0, 1, 2$) series we found an ubiquitous deviation from the idealized compositions~\cite{20, 27}. Henceforward, the title compound is denoted as Mn147, keeping in mind its non-stoichiometry, and MnBi$_2$Te$_4$ compound is denoted as Mn124.
 
 As shown first in~\cite{27}, Mn147 is thermodynamically stable in a high-temperature interval well above the room temperature. Whereas Mn124 melts at 600(5)~$^\circ$C~\cite{20}, the melting point of Mn147 is 590(5)~$^\circ$C~\cite{27} and, thus, offers a very narrow window above the crystallization point of Bi$_2$Te$_3$ (586~$^\circ$C), in which Mn147 can be grown from a melt. 
 Crystal-growth is thus very challenging, and our experiments evidence that polycrystals grown outside the determined temperature window exhibit stacking variants of both Mn124 and Mn147. Mn147 is not thermodynamically stable at room temperature, but can be obtained as a metastable product by  quenching from 585~$^\circ$C~\cite{27}. Based on these findings, we have established an optimized crystal-growth technique for Mn147: mm-sized platelets (Fig.~\ref{fig:Mn147struct}b) can be obtained by a stepwise slow cooling of a heterogeneous MnTe$_{(s)}$/Bi$_2$Te$_3$$_{(l)}$ melt and long-term annealing at a precisely controlled temperature of 585~$^\circ$C, followed by rapid water-quenching (see Appendix \ref{app:crystal} for further details). These high-quality crystals enable the following studies of the crystal structure, the magnetic order, the transport and the surface electronic structure (sections II and III). The compositions of all crystals used for physical property measurements were verified by energy-dispersive X-ray spectroscopy (EDX).

Our current structure elucidation by SCXRD confirms cationic non-stoichiometry in Mn147 in full accordance with our previous data on Mn147 powders~\cite{27}. The disorder manifests itself in Mn$^{2+}$/Bi$^{3+}$ antisite defects and related cationic vacancies ($\Box$) in the septuple (Mn$_{1-x}$$\Box_{x/3}$Bi$_{2+ 2x/3}$Te$_4$) blocks, predominantly in the \textit{1a} site in the middle of the septuple block (Tables S1, S2, Supplementary Note 1 \footnote{See Supplemental Material at [URL will be inserted by
publisher] for: Supplementary Notes 1-7, Figures S1-S13 and Tables S1-S4.}). Interestingly, the (Bi$_2$Te$_3$) block appears unaffected by cationic intermixing (Table S3). These mixed occupancies of the cationic positions and Mn vacancies result in a non-stoichiometric composition Mn$_{0.75(3)}$Bi$_{4.17(3)}$Te$_7$ as refined from a SCXRD experiment. This stoichiometry slightly deviates from the one previously determined for polycrystalline powders, Mn$_{0.85(3)}$Bi$_{4.10(2)}$Te$_7$~\cite{27}; thus, indicating that a homogeneity range $0.15 \leq x \leq 0.25$ may exist for Mn$_{1-x}$$\Box$$_{x/3}$Bi$_{4+{2x/3}}$Te$_7$ phase. 

The cationic disorder, however, does not alter the trigonal lattice symmetry of Mn147 (sp. gr. $P\bar{3}m$1; the GeBi$_4$Te$_7$ structure type~\cite{28}), neither inhibits long-range magnetic order. Similar intrinsic  phenomenon has been reported for isostructural~\cite{28}, and structurally~\cite{20} and compositionally related~\cite{29,30} compounds. In contrast to some of them, we find no indications of massive stacking faults in our crystals by X-ray or electron diffraction methods~\cite{27}. 

\section{Magnetic properties}

\begin{figure*}
  \center
  \includegraphics[width =\textwidth]{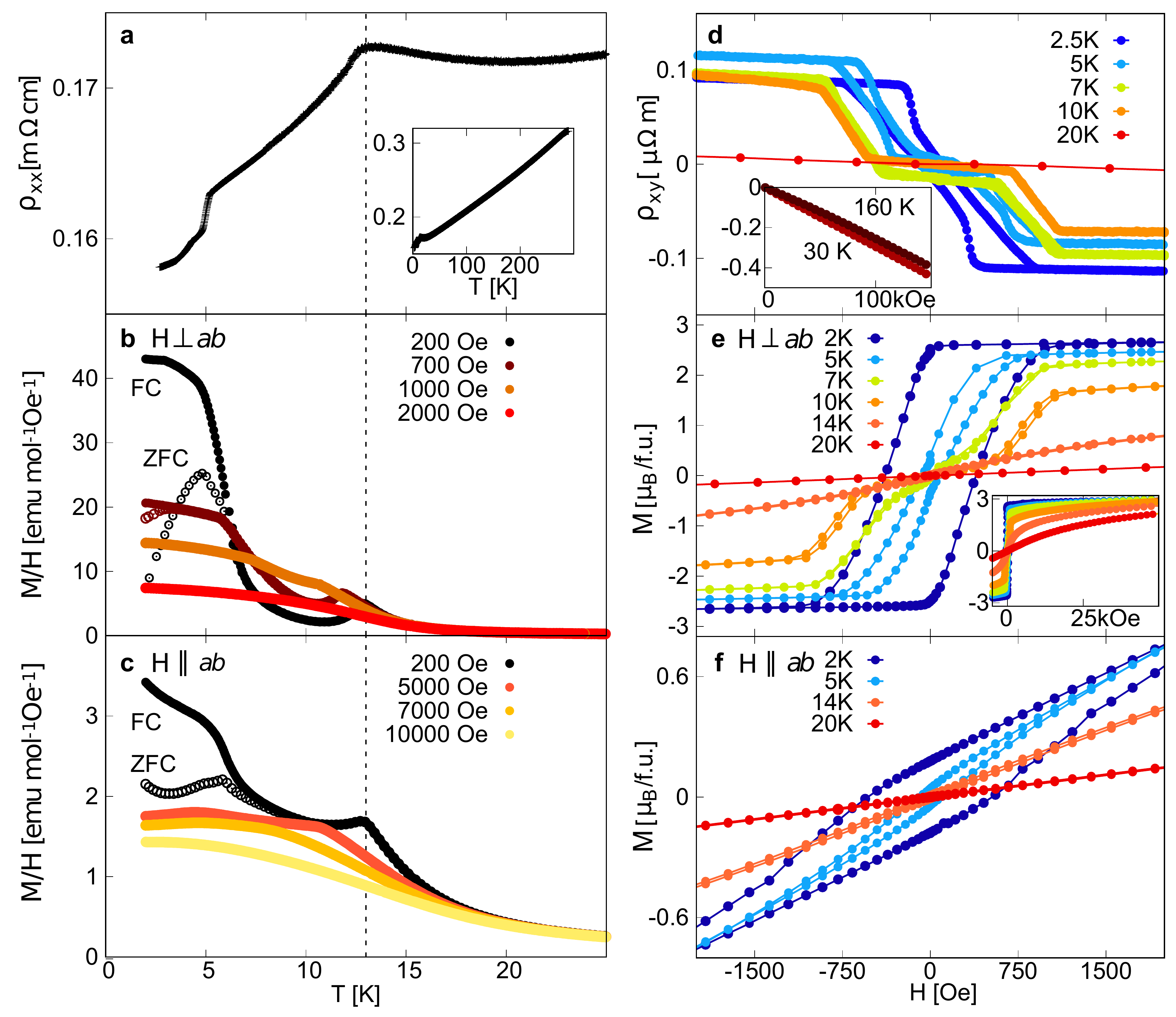}
  \caption{\label{fig:Mn147}  Magnetic and transport properties of Mn147.  \textbf{a}, In-plane electrical resistivity as a function of the temperature. \textbf{b} and \textbf{c}, Normalized magnetization as a function of temperature for fields applied both parallel and perpendicular to the $ab$ directions. Open and filled symbols correspond to ZFC and FC protocols, respectively. \textbf{d} and \textbf{e},  Hall resistivity and magnetization as a function of the field applied perpendicular to the $ab$ planes. \textbf{f} Magnetization as a function of the field applied parallel to the $ab$ planes. The hysteretic behavior for temperatures $T < 5$~K indicates  ferromagnetic (intra-plane) interactions.}
\end{figure*}

In this section, we analyze the bulk and surface magnetic properties of Mn147 crystals, based on which we discuss the topological electronic properties in section III. Electrical resistivity ($\rho_{xx}$) measurements as a function of temperature ($T$) reveal a metallic behaviour (see Fig.~\ref{fig:Mn147}a). Focusing on the most salient features of the data, a clear upturn anomaly is visible at 13~K, which is reminiscent of the typical signature of magnetic ordering in itinerant materials~\cite{31}. The upturn indicates enhanced fluctuations causing electron scattering, which is strongly reduced in the ordered phase, in which a steep decrease of $\rho_{xx}$ occurs. Upon lowering the temperature, a jump-like drop at about 5~K reveals further reduction of scattering, possibly related to a rearrangement of the magnetic structure.

\begin{figure*}
  \center
  \includegraphics[width = \textwidth]{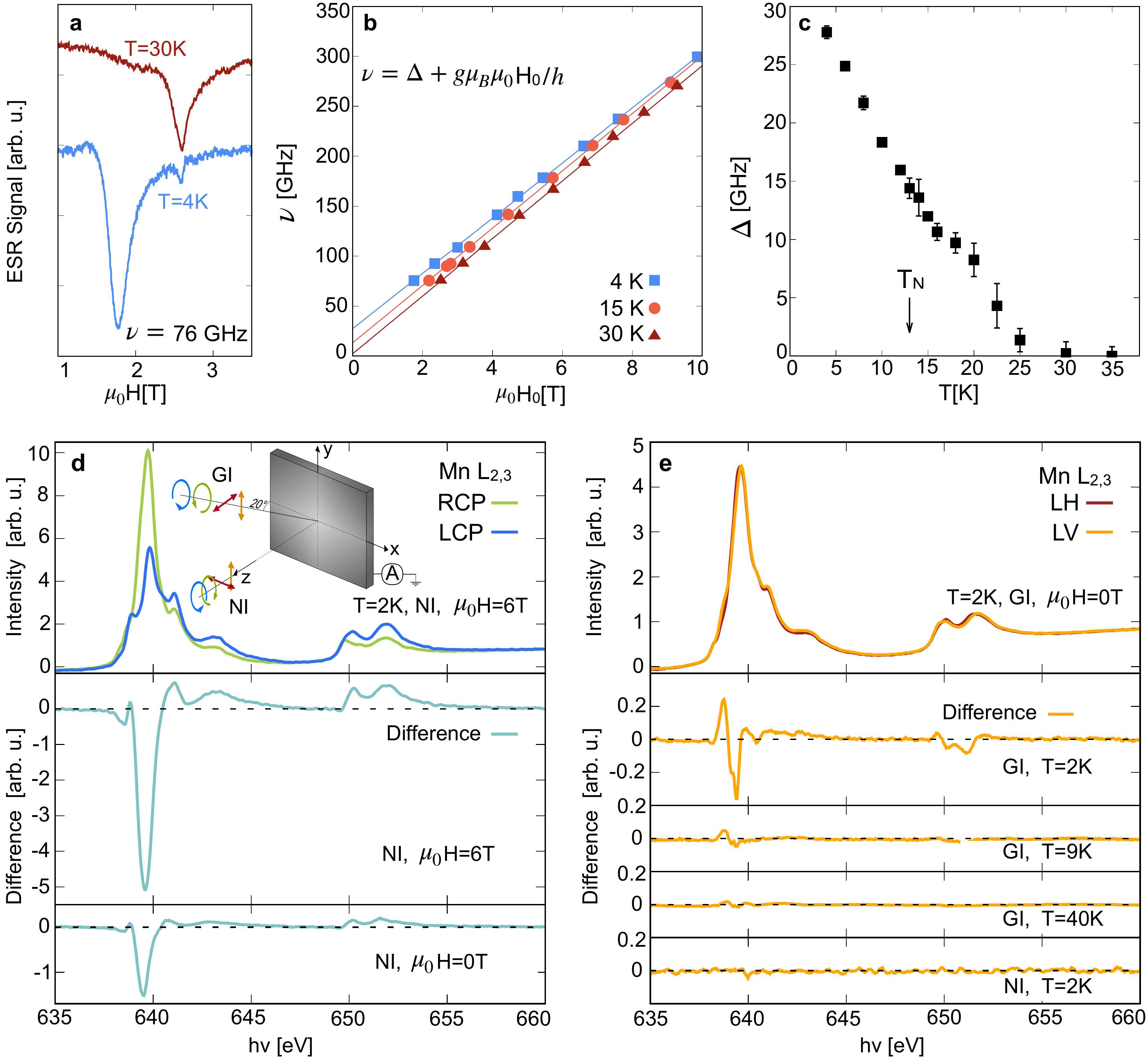}
  \caption{\label{fig:Mn147_2}  Spectroscopy of magnetic properties in Mn147. \textbf{a}, Typical ESR spectra measured at $T =$ 4~K and $T =$ 30~K. \textbf{b}, Frequency dependence of the resonance field of the ESR signal measured at temperatures of 4~K, 15~K, 30~K. \textbf{c}, The anisotropy gap $\Delta$ as a function of temperature extracted from \textbf{a} by fitting $\nu  = \Delta + g \mu_B \mu_0 H_0/ h$. \textbf{d}, XMCD and \textbf{e}, XMLD data for Mn147(0001) obtained at the Mn $L_{2,3}$ absorption edge with circularly polarized (RCP and LCP) and linearly polarized (LV and LH) light, respectively. Measurements were performed in normal (NI) and grazing (GI) light incidence geometries, as sketched in the inset of \textbf{d}. XMDC signals are shown for an external field ($\mu_0 H =$ 6~T) along the light incidence direction and for remnant conditions ($\mu_0 H = 0$) at $T =$ 2~K. XMLD data without external field are reported for different temperatures.}
\end{figure*}
Indeed, measurements of the magnetization ($M$) in an external field ($H$) on a Mn$_{0.82(7)}$Bi$_{4.2(1)}$Te$_{7.00(5)}$ crystal (as determined by EDX) as a function of temperature show an antiferromagnetic phase transition at $T_N =$ 13 K (Fig.~\ref{fig:Mn147}b). For $H \bot ab$ and small fields, such as $H =$ 200 Oe, a ferromagnetic-like increase occurs upon further cooling. This is followed by a splitting of field-cooled (FC) and zero field-cooled (ZFC) curves at around 7~K, as well as a kink and a peak at about 5~K, respectively, the latter coinciding with the jump in the resistivity. Both features are rapidly suppressed by applying an external magnetic field.

In the magnetically ordered phase, interesting metamagnetic behavior occurs for $H \bot ab$ as is evidenced by the magnetization curves (Fig.~\ref{fig:Mn147}e). For example, at $T =$ 10~K a spin-flip-like feature is observed, and for $T <$ 7~K dominating  hysteresis typical for ferromagnets is apparent. This complex magnetic phenomenology is also reproduced by the Hall resistivity $\rho_{xy}$ (Fig.~\ref{fig:Mn147}d). For $T < T_N$, the system exhibits anomalous Hall effect tracking the observed metamagnetic behavior. For $T \leq$ 7~K the data reveal an additional metamagnetic transition in the low-field region $\mu_0 H \leq$ 300~Oe. This observation is confirmed by a close inspection of the magnetization data. At 2.5~K a large hysteresis associated with global ferromagnetism is present. Above $T_N$ the anomalous contribution disappears, and only a standard component persists with a negative sign consistent with an $n$-type conduction (inset in Fig.~\ref{fig:Mn147}d).

The magnetic anisotropy of the compound is examined via additional measurements for $H \parallel ab$ (Fig.~\ref{fig:Mn147}c and f). The magnetic moment values in the ordered state are much lower in this case. At lower fields both an antiferromagnetic transition and a ZFC--FC splitting are observed, but the suppression of these features occurs at higher fields than for $H \bot ab$. For the $H \parallel ab$ direction the magnetization increases almost linearly with the applied magnetic field as expected for an antiferromagnet. On top of that, a spin reorientation at lower temperatures is indicated by an increase of $M/H$ below ca.~10~K and a ferromagnetic net magnetization is clearly present.  
Apparently, the field necessary to observe such feature can be sample-dependent, as is evident from a comparison of our  results with the recent reports \cite{Wueaax9989, PhysRevB.100.155144,arXiv:1910.11653}. This finding may be associated with slight differences in the Mn content due to Mn/Bi intermixing and Mn vacancies, which, however, have no influence on the lattice symmetry. Notably, all reports are in line regarding the behavior of magnetization as a function of magnetic field in both directions. The results show that the system is yet not fully saturated at $\mu_0 H =$ 5~T as indicated by a small slope at higher fields (inset in Fig.~\ref{fig:Mn147}e). High-field experiments are necessary to gain a better insight into the details of the magnetic phase diagram.

The Curie--Weiss fitting of the magnetization high-temperature data in both directions yields positive values of the Curie--Weiss temperature: $\theta_{CW}^{ab}=$ 13.7(5)~K and $\theta_{CW}^{c}=$ 14.7(5)~K, thus, confirming the predominantly ferromagnetic character of the largest (intra-plane) exchange interactions (see Supplementary Note~2, Fig.~S1 \cite{Note1}). In addition, the estimated effective magnetic moments, given the homogeneity range ($0.15 \leq x \leq 0.25$), fall into the ranges $5.2 \mu_B \leq \mu_{eff}^{ab} \leq 5.6 \mu_B$ and $5.1 \mu_B \leq \mu_{eff}^{c} \leq 5.5 \mu_B$, where $\mu_B$ is the Bohr magneton, which suggest the manganese(II) high-spin configuration $S = \frac{5}{2}$. The microscopic nature of the different magnetic states as a function of field and temperature is a matter of debate and requires further elucidation. 

The observed high-frequency electron spin resonance (ESR) signal of Mn147 (Fig.~\ref{fig:Mn147_2}a--c) is almost isotropic above $T \sim$ 30 K and follows a typical Mn(II)-ion paramagnetic resonance condition $h\nu = g \mu_B \mu_0 H_0 \mid m_s^z - (m_s^z \pm 1)\mid$ with the $g$-factor very close to 2. Here, $h$ is the Planck constant, and $m_s^z$ is the projection of the spin on the quantization (magnetic field) axis. Importantly, below $T \sim 30$\,K an energy gap $\Delta$ develops in the ESR response, and the resonance condition is modified to $\nu  = \Delta  + g \mu_B \mu_0 H_0/h$. The measured linear dependence of $\nu$ vs.\ $\mu_0 H_0$ for $\mu_0 H \parallel ab$ (Fig.~\ref{fig:Mn147_2}a) is typical for the wave vector $q = 0$ spin wave excitation (ferromagnetic resonance --- FMR) in an easy-axis-type ferromagnetically ordered material, where $\Delta$ represents the magnetic anisotropy gap. Considering the smallness of $\Delta$ as compared to the applied magnetic fields, such linearity is incompatible with the resonance response of an ordered collinear antiferromagnet in this field regime~\cite{32,33}. The opening of the excitation gap $\Delta$ at $T \leq$ 30 K and its gradual increase (Fig.~\ref{fig:Mn147_2}b) evidence significant ferromagnetic spin correlations on the time scale of ESR ($10^{-11}$~s) unrelated to 3D antiferromagnetic ordering which sets in at $T_N =$ 13~K. Therefore, given the pronounced low-dimensionality of the system, it is likely that the Mn-containing blocks are inherently ferromagnetic and give rise to a typical FMR signal in strong fields. At the same time, the application of a magnetic field suppresses the expected much weaker inter-layer antiferromagnetic coupling responsible for the 3D long-range order at $T_N$ in zero and small fields, whereas a paramagnetic state with strong intra-plane ferromagnetic correlations  --- denoted in the following as correlated paramagnet (CPM) ---  persists up to temperatures of the order 30 K.

To complement the magnetic characterization, we carried out X-ray magnetic circular (XMCD) and linear dichroism (XMLD) experiments at the Mn $L_{2,3}$ absorption edge in total electron yield mode (TEY) with a typical probing depth of a few nm~\cite{34}. The XMCD data collected at $T =$ 2~K provide evidence for a substantial remanent net magnetization of the Mn ions along the surface normal (Fig.~\ref{fig:Mn147_2}d), in sharp contrast to our previous observations for Mn124 with antiferromagnetic order~\cite{20}. This confirms that the spontaneous ferromagnetic polarization observed in the bulk magnetization data below $T \sim 5$~K extends up to the surface layers. A sizable XMLD signal in grazing light incidence and its absence in normal incidence further confirm the remnant out-of-plane magnetization (Fig.~\ref{fig:Mn147_2}e) in agreement with the ESR results. The XMLD signal gradually diminishes with increasing temperature, confirming its magnetic origin and indicating the transition into the paramagnetic regime, in line with our bulk magnetization results.

By density-functional calculations (DFT) we considered various possible magnetic structures for the ordered MnBi$_4$Te$_7$ model (see Supplementary Note~3 \cite{Note1}). The Mn atoms are found in the high-spin Mn(II) configuration, in agreement with the high-temperature magnetization measurements and similar to Mn124~\cite{20}. Our calculations show that the magnetic moments within the Mn layers prefer intra-plane ferromagnetic order with an out-of-plane spin configuration. The first-neighbor coupling is estimated as $-0.09$~meV$/\mu_B^2$, which is very close to the value reported for Mn124~\cite{19}. Furthermore, we find that antiferromagnetic ordering between the Mn layers (AFM1 state) results in a smaller total energy than the ferromagnetic ordering (FM state). The energy difference is, however, only about 0.5~meV/Mn atom, which is an order of magnitude smaller than in Mn124~\cite{18}. Moreover, this value is very close to the magnetic anisotropy energy, which yields about 0.5~meV/Mn atom in favor of the easy-axis configuration. These estimates corroborate a scenario with competing magnetic states differing slightly in energy, and, hence, a more complex magnetic response shown by Mn147, as compared to Mn124.

\section{Electronic structure and topological surface state}

Having established the crystal structure and the magnetic properties of Mn147, we will now discuss its electronic structure based on DFT calculations and angle-resolved photoemission (ARPES) experiments. The structural resemblance between Mn147 and Bi$_2$Te$_3$ opens the question, to what extent Mn147 inherits properties from the (Bi$_2$Te$_3$) building blocks. We begin our theoretical analysis of the topology of the electronic structure with an auxiliary calculation without spin polarization (see Supplementary Note~4~\cite{Note1}), which shows that, in the absence of magnetism, the system would be both a strong topological insulator and a topological crystalline insulator, just like Bi$_2$Te$_3$~\cite{35}.

\begin{figure}
  \center
  \includegraphics[width = 8.7cm]{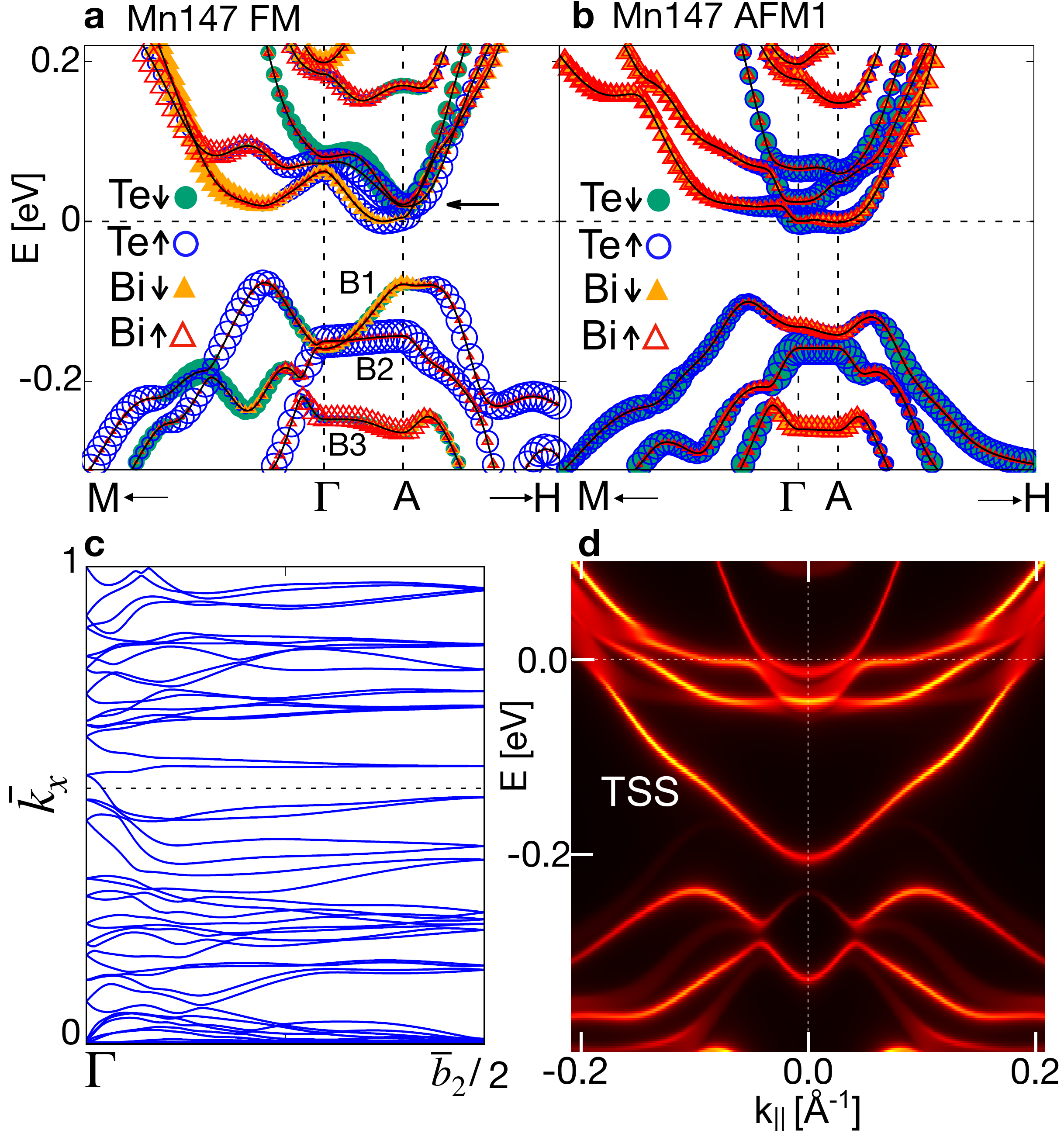}
  \caption{\label{fig:Mn147_3}  Band inversion phenomena in Mn147 (GGA+$U$+SOC). \textbf{a}, Band structure in the ferromagnetic configuration. The symbol size in each $k$-point and band is proportional to the overlap between the corresponding Bloch state and the Te and Bi $p$-orbitals, depicted in different colors. Filled (empty) dots correspond to spin down (up). The black arrow indicates the energy of the Weyl node of lowest energy in the conduction band. \textbf{b}, Band structure for the antiferromagnetic AFM1 configuration.
  \textbf{c}, Wannier center evolution in the $k_z=0$ plane. $k_x$ is the crystal momentum along the primitive lattice vector $\overline{b}_1$ and $\overline{b}_2$ is the second primitive vector in the $k_z =0$ plane. \textbf{d}, Mn147(0001) surface spectral density along the $\bar{\Gamma}-\bar{M}$ direction for a quintuple layer termination.}
\end{figure}

The influence of magnetism on the topological properties is first examined for the band-inversion phenomena. Fig.~\ref{fig:Mn147_3}a,b show the band structure of Mn147 for the FM and AFM1 (layer-wise AFM) order, respectively (see Fig.~S3 \cite{Note1}). The symbol size is proportional to the overlap between the corresponding Bloch states and the indicated orbitals. For reference, Fig.~S3f shows the well-known case of Bi$_2$Te$_3$, whose nontrivial topology originates in the inversion between Bi- and Te-$p_z$ orbitals of opposite parity~\cite{13,14,Note1}. The same orbitals contribute to the band inversion in Mn147 with the difference that the inverted bands are spin-polarized in the presence of a ferromagnetic component. Namely, the occupied Bi states form two bands (B1 and B3 in Fig.~\ref{fig:Mn147_3}a) of predominantly opposite spin. This effect is appreciable because the Bi atoms that are more involved in the band inversion are the ones closest to the Mn atomic layers. 

Naturally, in the AFM1 configuration the spin polarization in each band is compensated, i.e. each band is spin-degenerate. In this phase, the system realizes a ${Z}_2$ antiferromagnetic topological insulator (AFMTI), protected by a combination of time-reversal symmetry and translation along the $c$ axis. This case is analogous to the recently established AFMTI state in Mn124~\cite{19}. The ${Z}_2$ topology is analyzed in Fig.~\ref{fig:Mn147_3}c based on a Wannier charge center (WCC) in the $k_z=0$ plane. An arbitrary horizontal line crosses the WCC an odd number of times, and thus the $k_z=0$ plane behaves as a quantum spin Hall insulator \cite{PhysRevB.81.245209}. This ensures the existence of gapless surface states on side surfaces parallel to the $c$ axis. 

A natural question is to what extent the observed structure tendencies to Mn/Bi intermixing and to Mn vacancies affect the electronic structure.
To address this point we performed supercell calculations for the compositions \mnbi \, and \mnbiv \, (lifting up a restriction of electron neutrality used in the SCXRD refinements).
Our calculations show that the fundamental gap remains open, suggesting that, in the present amounts, Mn/Bi intermixing and Mn vacancies do not affect the non-trivial topology of the material, but rather only shift the chemical potential (see Supplementary Note~7~\cite{Note1}).

The existence of a topologically non-trivial surface state is confirmed by the calculated (0001) surface spectral density in Fig.~\ref{fig:Mn147_3}d for a quintuple-layer (QL) termination. As expected for the AFM1 state with an intra-plane ferromagnetic configuration the surface spectral density shows a gap-like feature at the $\bar{\Gamma}$-point \cite{19}. We find a similar situation in our surface calculations for the case of a septuple-layer (SL) termination (Fig. S8) and also for a FM magnetic configuration (Fig.~S5).

\begin{figure*}
  \center
  \includegraphics[width = \textwidth]{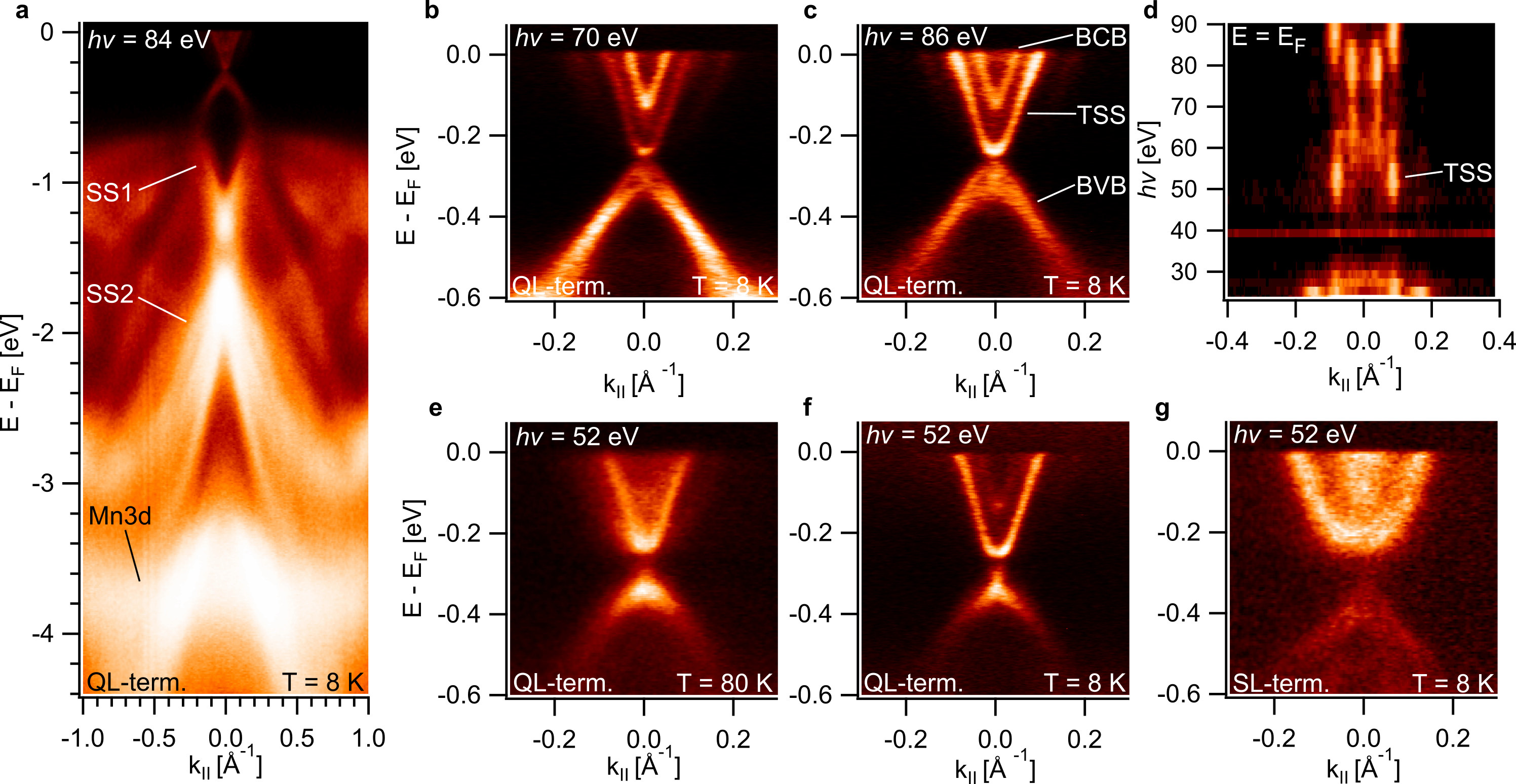}
  \caption{\label{fig:Mn147_4}  Electronic structure of the Mn147(0001) surface as measured by ARPES.  \textbf{a}, Overview data set of the valence band structure obtained at $T =$ 8~K showing characteristic surface states SS1 and SS2 and a feature related to Mn $3d$-states (cf. Fig.~S9--S12 \cite{Note1}). \textbf{b,c,f}, High-resolution data sets of the electronic structure near $E_F$ obtained at different photon energies and a temperature of $T =$ 8~K, showing a topological surface state (TSS) in the gap between conduction and valence-band derived states (BCB and BVB). \textbf{d}, Photon-energy dependence of the ARPES intensity at $E_F$ ($T =$ 8~K). \textbf{e}, Same as in \textbf{f}, but for $T =$ 80~K. \textbf{g}, Same as in \textbf{f}, but for a septuple layer (SL) termination. All other data sets in Fig.~4 are assigned to a quintuple layer (QL) termination. The ARPES data sets in Fig.~5 were measured along the $\bar{\Gamma} - \bar{M}$ high-symmetry direction.}
\end{figure*}

To experimentally support the non-trivial topology of the electronic structure predicted by our calculations we conducted ARPES measurements on the natural cleaving (0001) surface of Mn147 (Fig.~\ref{fig:Mn147_4}). The overview band structures (Fig.~\ref{fig:Mn147_4}a, Fig.~S9d,e in Supplementary Note~6 \cite{Note1}) bear clear resemblance to previous ARPES experiments for the topological insulator Bi$_2$Te$_3$~\cite{13,36}. Most importantly, we likewise find a state with a V-shaped dispersion in the bulk gap between the conduction and valence band states (BCB and BVB), near the Fermi level $E_F$ (Fig.~\ref{fig:Mn147_4}b,g). Systematic photon-energy-dependent measurements confirm the surface character of this state (Fig.~\ref{fig:Mn147_4}d). By comparison to our density-functional calculations in Fig.~\ref{fig:Mn147_3}d we identify it as a topologically non-trivial surface state (TSS). 

The observation of conduction band states below $E_F$ is in line with our transport measurements, although the prominent feature BCB shows a markedly 2D character and possibly arises from band bending, as commonly found for Bi$_2$Te$_3$~\cite{37} and Bi$_2$Se$_3$~\cite{38}. At higher binding energies in the valence band, we observe additional surface states SS1 and SS2 (Fig.~\ref{fig:Mn147_4}a, Fig.~S9d,e) that are similar to those previously detected for Bi$_2$Te$_3$~\cite{36}. These states are highly surface-localized well within a single (Bi$_2$Te$_3$) QL~\cite{36}, suggesting that our results in Fig.~\ref{fig:Mn147_4}a--f represent a surface terminated by a Bi$_2$Te$_3$ QL. This is supported by our calculations in Fig.~S7a \cite{Note1} for Bi$_2$Te$_3$-terminated Mn147, where similar surface states are found. Measurements on a single (0001) surface also revealed areas with a different well-defined band structure (Fig.~\ref{fig:Mn147_4}g, Fig.~S10), which we tentatively attribute to the second possible surface termination by a (MnBi$_2$Te$_4$) SL. The reduced data quality for this SL-termination may arise from the higher defect density in the SL than in the QL evidenced by our XRD results in section 1. Nevertheless, we observe qualitatively similar features in ARPES as for the QL-termination. Both terminations accommodate a dispersionless feature at a binding energy near 3.8~eV, which can be attributed to the Mn $3d$-states (Fig.~\ref{fig:Mn147_4}a, Fig.~S10), as confirmed by resonant photoemission measurements at the Mn $L$-edge (Fig.~S11).

Unlike for Bi$_2$Te$_3$, our measurements for Mn147 in the AFM1 state suggest the presence of a finite separation between the TSS and the BVB. This gap-like feature shows a subtle photon-energy-dependence arising mainly from changes in the spectral appearance of the BVB maximum, as exemplified by the three data sets in Fig.~\ref{fig:Mn147_4}b,c,f. The latter consists of at least two different features within a narrow energy range that exhibit complex $h\nu$-dependent intensity variations and possibly arise from a coexistence of surface- and bulk-derived states. Measurements at $T =$ 80~K (Fig.~\ref{fig:Mn147_4}e) do not show strong changes in the spectra for $h\nu =$~52 eV. However, towards higher temperatures we observe an increased spectral-weight filling of the gap-like feature, suggesting its mitigation with increasing temperature (Fig. S12). 

A comprehensive picture of the detailed spectral-weight behaviour of the TSS near the Dirac point is yet to emerge. Gap-like features, even in the paramagnetic regime, were also found in different magnetically doped TIs~\cite{39,40}, in Mn124 \cite{19,21} and, very recently, also in Mn147~\cite{Wueaax9989, arxiv:1910.13943, arXiv:1910.11653, arXiv:1905.02154}. We expect our detailed discussion
of the $h\nu$-dependency to be a useful ingredient, which, e.g., indicates a gap-like feature considerably smaller than the one observed in \cite{Wueaax9989}.
Additionally, other reports have found a vanishing gap in Mn147~\cite{arXiv:1910.11323, arXiv:1910.11014,PhysRevX.9.041039} and argued about a surface magnetic structure possibly different to that in the bulk. At low temperatures, our XMCD measurements provide evidence for a finite net magnetization at the surface, which motivates future ARPES measurements in this temperature regime. These developments shows that, likely, additional spectroscopic experiments, e.g. including spin-resolution, scanning tunneling microscopy and transport experiments on thin flakes will help to further elucidate this essential point.

\section{Magnetic topological phases}

\begin{figure}
  \center
  \includegraphics[width = 0.99\linewidth]{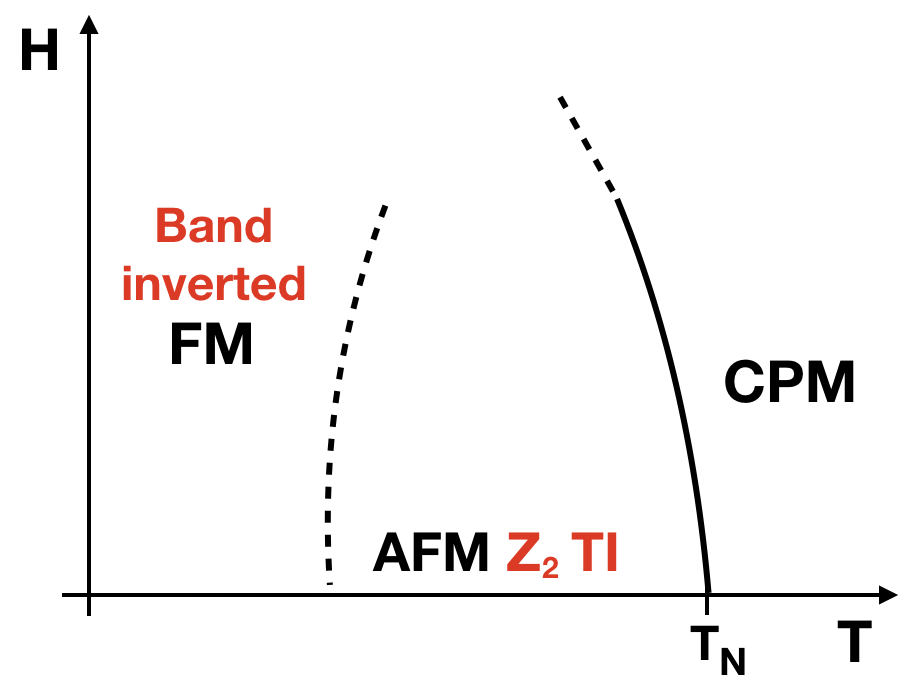}
  \caption{\label{fig:Mn147_5}  Schematic topological phase diagram of MnBi$_4$Te$_7$. The scheme follows the experimental observed trends: a correlated paramagnetic state above $T_N$, followed by an antiferromagnetic phase that at lower temperatures evolves to magnetic state with a strong ferromagnetic component. The text in red highlights possible non-trivial topology (see text for details). }
\end{figure}

Motivated by the experimental observations of a TSS due to band inversion and of the competition between different magnetic phases, we outline, based on our calculations, a topological phase diagram as a function of temperature and magnetic field. Fig.~\ref{fig:Mn147_5} sketches the phases theoretically explored in the ordered MnBi$_4$Te$_7$ model. Below the Ne$\acute{\text{e}}$l temperature, our calculations predict that the antiferromagnetic phase (AFM1) hosts a $Z_2$ topological insulating phase protected by a combination of the time-reversal and translation symmetries along the polar axis. In the lowest-temperature regime the experiments reveal a phase  with net magnetization at zero magnetic field (see Fig.~\ref{fig:Mn147} e and f). To mimic this regime we consider a collinear FM state which, interestingly, also features a non-trivial topology (see Supplementary Note~5 \cite{Note1}). Namely, this phase realizes a topological crystalline insulator (TCI) tunable by the magnetization orientation. Specifically, the crystal structure presents three mirror planes that contain the polar axis and are related by $2\pi /3$ rotations. When the magnetization points perpendicular to one of these planes, it preserves the corresponding reflection symmetry. The calculated mirror Chern number in such a magnetic configuration equals to $-1$. For other magnetization orientations, as in particular out-of-plane, the topological protection at the (0001) surface is lifted and the corresponding surface state is gapped (see Fig.~S5 \cite{Note1}). In addition, as shown in Fig.~\ref{fig:Mn147_3}a, doped samples can be of interest, since a ferromagnetic component splits the double degeneracy of the bulk bands and opens up the possibilities of Weyl physics. Indeed, Weyl nodes are revealed only 24~meV above the gap, very close to the bottom of the conduction bands (see the arrow in Fig.~\ref{fig:Mn147_3}a and Fig.~S6).

\section{Conclusion}

We have presented a comprehensive study of structural, magnetic and electronic properties of the Bi$_2$Te$_3$-derivative Mn$_{0.75(3)}$Bi$_{4.17(3)}$Te$_7$, i.e. the ($n = 1$)-member Mn147 of the modular (Bi$_2$Te$_3$)$_n$(MnBi$_2$Te$_4$) series. Our results indicate that Mn147 realizes an intrinsic magnetic topological insulator, similar to the recently established first antiferromagnetic topological insulator MnBi$_2$Te$_4$ for $n = 0$ \cite{19}. Unlike for MnBi$_2$Te$_4$, Mn147 develops a strong out-of-plane ferromagnetic component at low temperatures. In this regime Mn147 realizes the first instance of compound that features, both, an \textit{intrinsic} net magnetization and a topologically non-trivial surface state originating from a band inversion. In the thin-film limit these properties could facilitate the realization of the quantum anomalous Hall effect in an intrinsic material,  as recently reported for Mn124 where, however, it requires large external fields due to the robust antiferromagnetism~\cite{41,42}. 
Moreover, our calculations show how the complex magnetic phase diagram of Mn147, that we observe experimentally, may facilitate the tunability between different topological regimes, including antiferromagnetic topological and topological crystalline insulator states.

%\textit{Note added. } In the final stages of writing, two papers focusing on the magnetization, transport and photoemission in MnBi$_4$Te$_7$, although prepared by a different synthetic route, have appeared as arXiv:1905.02154 and arXiv:1905.02385.

\begin{acknowledgments}

We thank E.V.~Chulkov and M.M.~Otrokov (DIPC, San Sebastian, Spain, and Tomsk State University, Tomsk, Russia) for the initial impetus for this and other works on manganese-bismuth tellurides as perspective topological materials. This work was supported by the German Research Foundation (DFG) in the framework of the Special Priority Program (SPP 1666, IS 250/1-2) ''Topological Insulators'', by the ERA-Chemistry Program (RU 766/15-1), by CRC ''Tocotronics'' (SFB 1170), by CRC ''Correlated Magnetism -- From Frustration to Topology'' (SFB-1143, project id 247310070) and by W\"urzburg--Dresden Cluster of Excellence on Complexity and Topology in Quantum Matter -- \textit{ct.qmat} (EXC 2147, project-id 39085490). Part of this work was carried out with the support of the Diamond Light Source, beamline I05 (proposal SI22468-1).  We acknowledge experimental support by K.~Kissner, M.~\"Unzelmann, S.~Schatz, Chul Hee Min (University of W\"urzburg), F.~Diekmann, S.~Rohlf and M.~Kall\"ane (University Kiel) as well as the beamline staff at the Maestro endstation (ALS, Berkeley), at the APE beamline (Elettra, Trieste) and at the beamline P04 of PETRA III (DESY Hamburg). 
J.I.F. thanks the Alexander von Humboldt Foundation for financial support through the Georg Forster Research Fellowship Program.
K.~Mehlawat acknowledges the Hallwachs--R\"ontgen Postdoc Program of \textit{ct.qmat} for financial support.
J.I.F., R.R. and M.Ri. thank U. Nitzsche for technical assistance. J.I.F. and F.C. thank the IFW Excellence Program.

\end{acknowledgments}

%\subsection{Author Contributions}

%A.Z. synthesized and characterized the material assisted by A.I.; A.Z. and M.Ru. have elucidated the crystal structure; J.I.F. and R.R performed and interpreted the density-functional calculations with the help of M.R.; M.H.H., L.T.C.B. and A.U.B.W. conducted and analyzed magnetic field- and temperature-dependent magnetization measurements; F.C. and C.H. carried out and analyzed the transport measurements; K.M., A.A. and V.K. performed and analyzed the ESR experiments; R.C.V. conducted the photoemission and X-ray absorption spectroscopy studies and performed its analysis with assistance from H.B., S.M., T.F., T.R.F.P., H.B.V., M.V., S.J., C.C. The manuscript was written by H.B., J.I.F., A.I. and L.T.C.B. with contributions from all co-authors. A.I. and H.B. initiated and directed this research project. B.B., J.v.d.B., M.Ru. and F.R. supervised the project and contributed to discussions.

\section{Appendix}

In this Appendix we provide further information on the different methods used in this work.

\subsection{Crystal Growth}
\label{app:crystal}

First indications for the existence of Mn$_{1-x}$$\Box_{x/3}$Bi$_{4+2x/3}$Te$_7$ ($0.15 \leq x \leq 0.25$) and its composition were obtained in our previously published DSC experiments~\cite{20}. Attempts to synthesize a phase-pure powder of Mn$_{0.85}$Bi$_{4.10}$Te$_7$ following the synthetic route described in Ref.~\cite{20}, namely by long-term annealing of a stoichiometric mixture of Bi$_2$Te$_3$ and $\alpha$-MnTe at subsolidus temperature lead to considerable amounts (up to 15~wt.-\%) of MnBi$_2$Te$_4$ admixtures. A phase-pure ingot of Mn$_{0.85}$Bi$_{4.10}$Te$_7$ was synthesized by annealing at 590~$^\circ$C for 3~days, subsequent slow cooling to 585~$^\circ$C and, finally, annealing for 1~day followed by rapid quenching in water. Worth noting is that powders with the idealized MnBi$_4$Te$_7$ composition that were prepared by this route contained impurities, suggesting that this composition lies outside the homogeneity range. High-quality single crystals of Mn$_{1-x}$$\Box_{x/3}$Bi$_{4+2x/3}$Te$_7$ were grown by slow cooling ($-1$~K/h) of a melt from 650~$^\circ$C down to 585~$^\circ$C (right above the solidification point of Bi$_2$Te$_3$), followed by annealing for 10~days and rapid quenching. Platelet-like strongly intergrown crystals were mechanically extracted from the obtained ingots. Their compositions were controlled by EDX analysis.

\subsection{X-ray Diffraction and Energy-Dispersive X-ray Spectroscopy} 

Single-crystal X-ray diffraction data were collected on a four-circle Kappa APEX II CCD diffractometer (Bruker) with a graphite(002)-monochromator and a CCD-detector at $T =$ 296(1)~K. Mo-K$_\alpha$ radiation ($\lambda =$ 71.073~pm) was used. A numerical absorption correction based on an optimized crystal description was applied~\cite{45}, and the initial structure solution was performed in JANA2006~\cite{46}. The structure was refined in SHELXL against $F_o^2$~\cite{47}. Further details on the crystal structure investigations of Mn$_{0.75(3)}$Bi$_{4.17(3)}$Te$_7$ can be obtained from the Fachinformationszentrum Karlsruhe, 76344 Eggenstein-Leopoldshafen, Germany (fax, (+49)7247-808-666; E-mail, crysdata@fiz-karlsruhe.de), on quoting the depository number CSD-1891486.

Powder X-ray diffraction data were measured using an X-Pert Pro diffractometer (PANalytical) with Bragg-Brentano geometry or a Huber G670 diffractometer with an integrated imaging plate detector and read-out system. Both machines operate with a curved Ge(111) monochromator and Cu-K$_{\alpha 1}$ radiation ($\lambda =$ 154.06~pm). Variable divergence slits were used on the X-Pert Pro equipment to keep the illuminated sample area constant. The graphics of the structures were developed with the Diamond software~\cite{48}.

Energy dispersive X-ray spectra (EDX) were collected on a scanning electron microscope Hitachi SU8020 using an Oxford Silicon Drift X-MaxN detector at an acceleration voltage of 20~kV and 100~s accumulation time. The EDX analysis was performed using the $P$/$B$-$ZAF$ standardless method (where $Z$ = atomic no. correction factor, $A$ = absorption correction factor, $F$ = fluorescence factor, and $P$/$B$ = peak to background model). Experimentally determined compositions (EDX) fall into a range from Mn$_{0.6(1)}$Bi$_{4.4(1)}$Te$_7$ to Mn$_{0.8(1)}$Bi$_{4.3(1)}$Te$_7$.

\subsection{Angle-Resolved Photoelectron Spectroscopy}

ARPES measurements on the (0001) surface of cleaved crystals in a temperature range range between 8~K and 80~K were carried out at the high-resolution-branch of beamline i05 at the Diamond Light Source, UK, using p-polarized photons with energies between $h \nu =$ 20 and 90~eV and energy resolution $< 10$~meV [Fig. 5 and S10]. The spot size of the photon beam was ca. 30 $\mu$m.   Supplementary ARPES experiments were performed at the Microscopic and Electronic Structure Observatory (MAESTRO) at beamline~7 of the Advanced Light Source (ALS) [Fig. S9] at the LE-branch of APE beamline at the Elettrasynchrotron [Fig. S11]. All measurements were performed in ultrahigh vacuum of lower than $10^{-10}$~mbar. Supplementary core-level photoemission data were acquired at the ASPHERE III endstation at beamline P04 of PETRA III (DESY, Hamburg) [Fig. S11].

\subsection{Electron Spin Resonance Measurements}

ESR experiments were performed on a single crystal with a home-made ESR setup in the microwave frequency range $\nu =$ 75--300~GHz, in the temperature range $T =$ 4--35~K and in magnetic fields up to $\mu_0 H_0 =$ 16~T.

\subsection{X-ray Magnetic Circular and Linear Dichroism}

XMCD and XMLD measurements on the (0001) surface of cleaved crystals were carried out in total electron yield (TEY) mode at the BOREAS beamline of the ALBA synchrotron \cite{xmcd}.

\subsection{Density-Functional Calculations}

Fully-relativistic Density Functional Theory (DFT) calculations were performed using the PBE implementation~\cite{49} of the Generalized Gradient Approximation (GGA) and treating the spin-orbit coupling in the 4-spinor formalism, as implemented in FPLO-18~\cite{50}. For results presented in the main text, the experimental crystal structure based on a fully ordered MnBi$_4$Te$_7$ was used in our calculations. Namely, the cationic intermixing and cation deficiency were neglected and the stoichiometric limit MnBi$_4$Te$_7$ was considered.
Effects of these sorts of defects were studied by supercell calculations (see Supplementary Note 7 \cite{Note1}).
For the ordered model calculations, a linear tetrahedron method with a mesh of $16 \times 16 \times 2$ subdivisions (or $16 \times 16 \times 1$ in the AFM1 state) in the full Brillouin zone was used. GGA+$U$ calculations we also performed using the atomic limit implementation of the double-counting correction and fixing $J = 1$~eV. The value of $U$ affects the resulting bulk gap and determines at which energies the spectral weight associated with Mn-$d$ states is placed. We find that the position of the Mn $3d$ states measured with core-level spectroscopy (see Fig.~S11 \cite{Note1}) is best described by a moderate value of $U \sim 2$~eV (see Fig.~S3c \cite{Note1}). This value renders a bulk gap of $\sim 75$~meV. In MnBi$_2$Te$_4$, higher values of $U$ have been used aiming to reproduce the experimental estimation of the gap~\cite{19}. The difficulty in finding a single value of $U$ that correctly accounts for all experimental results suggests that a quantitative comparison in these materials may necessitate the usage of exchange and correlation functionals beyond GGA+$U$. The statements on the total energy calculations are, however, robust, as shown in Supplementary Note~3. For the surface spectral calculations, as well as for the search of Weyl nodes, an accurate tight-binding model was built by constructing Wannier functions with the projection method implemented in the PYFPLO interface of FPLO~\cite{50}. The Bi-$6p$, Te-$5p$ and Mn-$3d$ orbitals were considered in this construction. The mirror Chern numbers were computed based on this Hamiltonian as implemented in Ref.~\cite{PhysRevMaterials.3.074202}.

\vspace{2mm}
\subsection{Magnetization and Transport}

The dc magnetization measurements were performed in a Superconducting Quantum Interference Device (SQUID) Vibrating Sample Magnetometer (VSM) from Quantum Design for 1.8~K up to room temperature and in magnetic fields up to 7~T. Zero-field-cooled (ZFC) and field-cooled (FC) magnetization curves were recorded upon warming. %The dc magnetization data was corrected for demagnetization effects.

The transport properties were performed in the standard 4-wires configuration using a home-made probe inserted in a He-bath cryostat by Oxford Instruments, endowed with a 15/17~T magnet.

\bibliography{references}

\end{document}